\documentclass[preprint,groupedaddress,pre,onecolumn]{revtex4}
\usepackage{amssymb}
\usepackage{amsmath}

\input{tcilatex}

\begin{document}

\title{Regularization of the collision in the electromagnetic two-body
problem}
\author{Efrain Buksman Hollander and Jayme De Luca}
\email[corresponding author; email address:]{ deluca@df.ufscar.br}
\affiliation{Universidade Federal de S\~{a}o Carlos, \\
Departamento de F\'{\i}sica\\
Rodovia Washington Luis, km 235\\
Caixa Postal 676, S\~{a}o Carlos, S\~{a}o Paulo 13565-905}
\date{\today }

\begin{abstract}
We derive a differential equation that is regular at the collision of two
equal-mass bodies with attractive interaction in the relativistic
action-at-a-distance electrodynamics. We use the energy constant related to
the Poincar\'{e} invariance of the theory to define finite variables with
finite derivatives at the collision. The collision orbits are calculated
numerically using the regular equation adapted in a self-consistent
minimization method (a stable numerical method that chooses only nonrunaway
solutions). This dynamical system appeared 100 years ago as an example of
covariant time-symmetric two-body dynamics and aquired the status of
electrodynamics in the 1940's by the works of Dirac, Wheeler and Feynman. We
outline the method with an emphasis on the physics of this complex
conservative dynamical system.
\end{abstract}

\pacs{05.45.-a}
\maketitle

\section{Introduction}

\bigskip

\textbf{Delay equations play an essential part in Maxwell's electrodynamics
because of the finite speed of propagation of the electromagnetic fields.
Another more modern version of electrodynamics, action-at-a-distance
electrodynamics\cite{Fey-Whe}, developed in the 1940's as an alternative to
avoid the divergencies of perturbative quantum electrodynamics\cite%
{Fey-Whe,Mehra}, was stopped by a main difficulty: delay equations. The
fundamental problem of a non-perturbative calculation of the level shifts of
hydrogen depends on our ability to deal with delay equations}. \textbf{Given
a Hamiltonian description for this electromagnetic two-body problem\cite{two}%
, knowledge of the orbits can be used} \textbf{in EBK quantization}. \textbf{%
Besides delay, an extra difficulty with the numerical calculation of an
orbit for the relativistic two-body problem is the collision, where the
equations of motion in usual form become singular. This obstacle has so far
prevented the numerical study of this problem. In this work we derive a
delay differential equation that is regular at the collision of two
equal-mass bodies with attractive interaction in the action-at-a-distance
electrodynamics. The collision orbit is calculated numerically with a
self-consistent minimization method that integrates the regularized equation 
\cite{two}.\ We calculate numerically four collision orbits with energies
from the nuclear to the atomic range.}

The usual Hamiltonian description of two-body dynamics is surprisingly
restrictive within relativity physics: If Lorentz transformations are to be
represented by canonical transformations, only non-interacting two-body
motion can be described. This is the content of the no-interaction theorem
of 1964, which later in 1984 was proved for the local Lagrangian description
as well \cite{nointeract}. A covariant version of Hamiltonian dynamics,
constraint dynamics, was invented to overcome this group-theoretical
obstacle, but it has a limited applicability \cite{Comar}, and in particular
a constraint description of electrodynamics is not known at present. In the
light of \cite{nointeract}, the only available description of
Lorentz-invariant two-body dynamics is via a Lagrangian built from the
scalar invariant of the Poincar\'{e} group, (the modulus of the separation
four-vector), that involves the time coordinates and yields delay equations
of motion. This is a second, more modern confirmation that delay equations
are in relativity physics and electrodynamics to stay. The idea to remove
the field degrees of freedom from electrodynamics goes back to Dirac \cite%
{Dirac} and later Wheeler and Feynman understood in 1945 that
action-at-a-distance electrodynamics was a theory that did not need
renormalization. The subsequent program to quantize the two-body problem of
the action-at-a-distance electrodynamics faced mathematical difficulties
that can be summarized in one word: delay equations. History says that the
famous seminar that never came from Wheeler (see Ref. \cite{Mehra}, page 97)
was due to difficulties in dealing with delay equations. In this same
chapter 5, page 97 of reference \ \cite{Mehra}, Feynman says that \ ` I didn
't solve it either---a quantum theory of half-advanced half-retarded
potentials---and I worked on it for years... '. Since then this has been an
open problem of atomic physics, with the main subsequent inputs coming from
the works of applied mathematicians\cite{Drivergroup}.

In 1903, Schwarzchild proposed a relativistic type of interaction between
charges that was time reversible precisely because it involved retarded and
advanced interactions symmetrically \cite{Schwarz}. The same model
reappeared in the 1920s in the work of Tetrode and Fokker \cite%
{Tetrode-Fokker} and it finally became an interesting physical theory after
Wheeler and Feynman showed that this direct-interaction theory can describe
all the classical electromagnetic phenomena (i.e. the classical laws of
Coulomb, Faraday, Amp\`{e}re, and Biot-Savart) \cite{Fey-Whe,Leiter}.
Wheeler and Feynman also showed in 1945 that in the limit where the electron
interacts with a completely absorbing universe, the response of this
universe to the electron's field is equivalent to the \emph{local}
Lorentz-Dirac self-interaction theory \cite{Dirac} without the need of mass
renormalization \cite{Fey-Whe,Narlikar}. \ The Wheeler and Feynman program 
\cite{Mehra} to quantize the action-at-distance electrodynamics and overcome
the infinities of QED is still not implemented because of the lack of a
Hamiltonian description\cite{two}. As very little is known of this important
physical problem at an analytical level, the knowledge of the trajectories
can be useful in EBK quantization\cite{two}.

The isolated two-body system, away from the other charges of the universe is
a conservative time-reversible dynamical system in the action-at-a-distance
electrodynamics. We consider here only the equal-mass two-body system ($%
m_{1}=m_{2}=m$), henceforth called 1D-WF2B. The only postulate of the
relativistic action-at-a-distance electrodynamics is that equations of
motion be derived formally \cite{Staruskiewicz} by extremizing the
parametrization-independent action 
\begin{equation}
S_{F}=-\int mds_{1}-\int mds_{2}-e_{1}e_{2}\int \int \delta
(||x_{1}-x_{2}||^{2})\dot{x}_{1}\cdot \dot{x}_{2}ds_{1}ds_{2},
\label{Fokker}
\end{equation}%
where $x_{i}(s_{i})$ represents the four-position of particle $i=1,2$
parametrized by its arc-length $s_{i}\,$, double bars stand for the
four-vector modulus $||x_{1}-x_{2}||^{2}\equiv (x_{1}-x_{2})\cdot
(x_{1}-x_{2})$, and the dot indicates the Minkowski scalar product of
four-vectors with the metric tensor $g_{\mu \nu }$\ ($%
g_{00}=1,g_{11}=g_{22}=g_{33}=-1$) (the speed of light is $c=1$). The
attractive problem is defined by Eq.(\ref{Fokker}) with $e_{1}=-e_{2}\equiv
e $ (positronium atom), while the repulsive two-electron problem is defined
by Eq. (\ref{Fokker}) with $e_{1}=e_{2}\equiv e$. For the repulsive
two-electron problem along symmetric orbits [$-x_{2}(t)=x_{1}(t)\equiv x(t)$%
], minimization of action (\ref{Fokker}) prescribes the following equation
of motion 
\begin{equation}
m\frac{d}{dt}(\frac{v}{\sqrt{1-v^{2}}})=\frac{e^{2}}{2r^{2}}\left( \frac{%
1-v(t-r)}{1+v(t-r)}\right) +\frac{e^{2}}{2q^{2}}\left( \frac{1+v(t+q)}{%
1-v(t+q)}\right) ,  \label{repulsemotion}
\end{equation}%
where $v(t)\equiv dx/dt$ is the velocity of the first electron, of mass $m$
and charge $e$, and $r$ and $q$ are the time-dependent delay and advance,
respectively. The functions $r(t)$ and $s(t)$ are implicitly defined by the
light-cone conditions%
\begin{eqnarray}
r(t) &=&x(t)+x(t-r),  \label{lightcone} \\
q(t) &=&x(t)+x(t+q).  \notag
\end{eqnarray}%
In general, a neutral-delay equation such as Eqs. (\ref{repulsemotion}) and (%
\ref{lightcone}) requires a pair of world-line segments of trajectory as the
initial condition (one world-line segment for each particle). As discussed
in Ref. \cite{wellposed}, the initial world-line segments can be provided in
such a way that Eqs.(\ref{repulsemotion}) and (\ref{lightcone}) are
well-posed, by using "maximal independent segments ". A pair of world-line
segments is called independent \ if the end points of each segment lie on
the forward and backward light-cones of a single point interior to the other
segment. \ Last, a surprising existence theorem was proved for the symmetric
motion of two electrons along a straight line [$-x_{2}(t)=x_{1}(t)\equiv
x(t) $] (Eqs. (\ref{repulsemotion}) and (\ref{lightcone}) ). For this simple
motion and for sufficiently low energies, it was shown in Ref. \cite%
{Drivergroup}, that Newtonian initial conditions [ $x(0)=x_{o}$ and $%
v(0)=v_{o}$ ] determine the unique solution that is globally defined (i.e.,
that does not runaway at some point) \cite{Drivergroup}.
Existence/uniqueness proofs are still lacking for the case of attractive
interaction, and we hope that with the present regularization of the
equations of motion such proofs can be facilitated.

For the relativistic two-body system, the only known analytical solution is
the circular orbit for the attractive problem \cite{Schonberg,Schild}. The
first numerical method to solve Eqs. (\ref{repulsemotion}) and (\ref%
{lightcone}) was developed in \cite{VonBaeyer} and converged to solutions up
to $v/c=0.94.$ Later another method \cite{Igor-pair}\ converged up to $%
v/c=0.99$. In reference \cite{two} we developed a numerical method for the
repulsive problem. Precisely because of the singularity, numerical methods
and studies for the attractive problem are lacking, and the method developed
in \cite{Bonn} has no hope of dealing with a near-collision. We hope that
this work can start to fill this gap. This paper is organized as follows: In
Section II we develop familiarity with the collision orbit and
regularization issues. In Section III we study the behavior of the symmetric
orbit near the collision with formal series expansions, and motivate the
change of the evolution parameter (the time transformation). As this alone
is not enough to accomplish manifest regularization, in Section IV we
introduce the energy constants of the electromagnetic two-body problem to
aid in the definition of two finite variables with manifestly finite
derivatives (i.e. our regular differential equation). Because of the delay
nature of the equations, coordinate transformations alone are not enough to
prove that the derivatives are finite at the collision. In appendix A we
make use of the energy constants to recognize the mathematical space for
regular orbits and to prove the regularity of these derivatives. The
material of appendix A provides an elegant alternative to the pedestrian
construction of formal power series in the neighborhood of the collision
(i.e., the material of section III). Last, in Section V we adapt the regular
equation in a numerical method that integrates future and past histories
until the eventual self-consistency of the histories. In this section we
also calculate numerically several orbits in several energy ranges, and the
implementation needed in each range to speed up convergence of the method.
In section VI we put the conclusion and discussions.

\section{The equations of motion}

Our regularization follows closely the Levi-Civita regularization of the
Galilei-invariant Kepler problem \cite{Tulio,Stiefel,Aarseth}, with the
additional difficulties imposed by the Poincar\'{e} invariance (i.e. delay).
As with the Levi-Civita regularization, a time transformation alone does not
accomplish regularization, and it is necessary to use the energy constant to
remove infinities from the equation of motion. In the present Poincar\'{e}%
-invariant case, besides a time transformation and use of the energy
constant, it is further necessary to define special finite variables to
accomplish manifest regularization. The non-local expression for the
conserved energy of the electromagnetic two-body problem \cite%
{Fey-Whe,Anderson,Klimenko} is therefore used here in two ways: (i) As in
the Levi-Civita regularization, to remove infinities from the equation of
motion and (ii) to define the required finite variables with manifestly
finite derivatives and to aid in the proof of regularity of these
derivatives. Unlike the Levi-Civita regularization,\textbf{\ }because of the
delay nature of the equations, it is not possible to check the regularity
just by performing coordinate transformations and taking limits, and the use
of the energy expression provides an elegant way to perform these limits and
to recognize the correct space of definition of the regular orbits, as
discussed in appendix A.\textbf{\ }With that we accomplish manifest
regularization.

\textbf{\ }We henceforth use a unit system where $m=c=e_{1}=-e_{2}=1$. We
assume that at $t=-t_{C}$ \ particle $1$ is at $x_{1}=0$ and moving to the
right while particle $2$ is at the same point and moving to the left
(outgoing collision). The particles collide again at $t=t_{C}$ \ with
ingoing velocities, so that $x_{1}\left( t\right) >0,$ $x_{2a}\left(
t\right) $ and $x_{2b}\left( t\right) <0$ all along the unit cell of our
orbit (see Figure 1). Because the transformations involved in the
regularization are elaborate, we choose to work in the special Lorentz frame
where the orbit is symmetric and therefore loose covariance in benefit of
the intuitive picture. A covariant analysis shall be left for later work. In
this work we consider only symmetric orbits of the equal-mass attractive
1D-WF2B, whose equation of motion for particle $1$ is

\begin{equation}
\frac{dv_{1}}{dt_{1}}=-\frac{1}{2}\left\{ \frac{1}{r_{a}^{2}}\frac{\left(
1-v_{2a}\right) }{\left( 1+v_{2a}\right) }+\frac{1}{r_{b}^{2}}\frac{\left(
1+v_{2b}\right) }{\left( 1-v_{2b}\right) }\right\} \left( 1-v_{1}^{2}\right)
^{3/2},  \label{eq1}
\end{equation}%
while the condition of symmetric colinear motion defines the trajectory of
particles $2$ as

\begin{equation}
x_{2}\left( t\right) =-x_{1}\left( t\right) .  \label{simetric/orbit}
\end{equation}%
In Eq.(\ref{eq1}) $v_{1\text{ }}$stands for the instantaneous velocity of
particle $1$ (present time is the time $t_{1}$ of particle $1$), while $%
v_{2b}$ and $v_{2a}\ $stand for the velocities of particle $2$ at the
retarded and advanced light-cones, respectively. As with Eqs.(\ref%
{repulsemotion}) and (\ref{lightcone}), the electrodynamic interaction in
Eq. (\ref{eq1}) connects points that are in light-cone condition, as defined
by 
\begin{eqnarray}
r_{a} &\equiv &\left\vert x_{1}\left( t_{1}\right) -x_{2}\left(
t_{2a}\right) \right\vert =c\left( t_{2a}-t_{1}\right) ,  \label{light-cone2}
\\
r_{b} &\equiv &\left\vert x_{1}\left( t_{1}\right) -x_{2}\left(
t_{2b}\right) \right\vert =c\left( t_{1}-t_{2b}\right) .  \notag
\end{eqnarray}%
In all the above, subscripts $a$ and $b$ indicate future and past times of
particle $2$ in light-cone to the present time $t_{1}$ of particle 1 (see
Figure 2).\ By use of Eq. (\ref{light-cone2}), we can also express $t_{2a}$
and$\ t_{2b\ }\ $as 
\begin{eqnarray}
t_{2a} &=&t_{1}+r_{a}/c,  \label{deft} \\
\ t_{2b} &=&t_{1}-r_{b}/c,  \notag
\end{eqnarray}%
and from Eqs.(\ref{light-cone2}) and (\ref{deft}) we can derive the equation
of motion for $t_{2a}$%
\begin{equation}
\frac{dt_{2a}}{dt_{1}}=\frac{\left( 1+v_{1}\right) }{\left( 1+v_{2a}\right) }%
.  \label{neeq2}
\end{equation}%
In the same way \ that we derived Eq. (\ref{neeq2}), the motion of $x_{2a}$, 
$x_{2b}$, and $t_{2b}$ can also be derived from Eqs. (\ref{light-cone2}) and
(\ref{deft}) \ as 
\begin{eqnarray}
\frac{dx_{2a}}{dt_{1}} &=&v_{2a}\frac{\left( 1+v_{1}\right) }{\left(
1+v_{2a}\right) },  \label{neeq3} \\
\frac{dx_{2b}}{dt_{1}} &=&v_{2b}\frac{\left( 1-v_{1}\right) }{\left(
1-v_{2b}\right) }.  \notag \\
\frac{dt_{2b}}{dt_{1}} &=&\frac{\left( 1-v_{1}\right) }{\left(
1-v_{2b}\right) },  \notag
\end{eqnarray}%
Eqs. (\ref{eq1}) and (\ref{neeq2})-(\ref{neeq3}),\ together with the
definitions of $r_{a}$ and $r_{b}$\ of Eq. (\ref{light-cone2}) constitute
the complete delay equation that we consider in this paper. Along a
symmetric orbit ( as defined by Eq. (\ref{simetric/orbit})), the retarded
and advanced velocities $v_{2b}$ and $v_{2a}$ are defined from the velocity $%
v_{1}(t)$ of particle $1$ by\ 

\begin{eqnarray}
v_{2a}\left( t\right) &\equiv &-v_{1}\left( t+r_{a}/c\right) ,
\label{defretadv} \\
v_{2b}\left( t\right) &\equiv &-v_{1}\left( t-r_{b}/c\right) ,  \notag
\end{eqnarray}%
as illustrated in Figure 2. By use of Eqs.(\ref{simetric/orbit}) and (\ref%
{light-cone2}), we can also show that along a symmetric orbit the two
quantities $r_{a}\left( t_{1}\right) \;$and\ $r_{b}\left( t_{1}\right) \;$%
are defined by a single function $\ r\left( t\right) $ as%
\begin{eqnarray}
r_{a}\left( t\right) &=&r\left( t\right) ,  \label{familia_r} \\
r_{b}\left( t+r_{a}/c\right) &=&r\left( t\right) ,  \notag
\end{eqnarray}%
which is illustrated in Fig. 3. As the force is always attractive from both
retarded and advanced positions in Eq. (\ref{eq1}), after the velocities
have switched opposite the interparticle distance must approach zero until
the collision happens. One could conjecture of orbits where the particles
reach the speed of light even before the collision. Such orbits, if they
exist, will not be studied here. We shall show below that both velocities
must tend to the speed of light as the particles approach the collision.

To prove that the velocities must go to the speed of light at the collision,
we need some monotonicity properties: From Eq.(\ref{eq1}) it follows that $%
\frac{dv_{1}}{dt_{1}}<0$ for all times, and by symmetry one also has $\frac{%
dv_{2a}}{dt_{2a}}>0\;$and $\frac{dv_{2b}}{dt_{2b}}>0$ , such that we can
establish that 
\begin{equation}
\left\vert v_{2a}\right\vert <v_{1}<\left\vert v_{2b}\right\vert ,
\label{monot_v}
\end{equation}%
for any time $-t_{C}<t_{1}<t_{C}.$ It is also easy to show that $r_{a}\left(
t_{1}\right) $ and $r_{b}\left( t_{1}\right) $ are piecewise monotonic
functions of $t_{1}.\;$From Eqs. (\ref{light-cone2}) and (\ref{neeq3}) it
follows that

\begin{equation}
\frac{dr_{a}}{dt_{1}}=\frac{v_{1}-v_{2a}}{1+v_{2a}},  \label{eq7}
\end{equation}%
and\ 
\begin{equation}
\frac{dr_{b}}{dt_{1}}=\frac{v_{1}-v_{2b}}{1-v_{2b}}.  \label{eq8}
\end{equation}%
As the velocities are globally monotonic, there is a maximum radius $r_{0}\ $%
(see Figure 1) attained at $t_{1}=t_{0}>-t_{C}\ ,$when $v_{1}\left(
t_{1}=t_{0}\right) =v_{2a}\left( t_{2a}\right) $, (such that $\left. \frac{%
dr_{a}}{dt_{1}}\right\vert _{t_{1}=t_{0}}=0)$. It must be that 
\begin{eqnarray}
\frac{dr_{a}}{dt_{1}} &>&0\ \text{;\ }\ t_{1}<t_{0},  \label{monot_r} \\
\ \frac{dr_{a}}{dt} &<&0\ \text{;\ \ }\ t_{1}>t_{0}.  \notag
\end{eqnarray}%
As the collision happens at $t_{1}=-t_{C}<t_{0},$ we can restrict to the
increasing part of $r_{a},$ $(-t_{C}<t<t_{0})$ , an interval where Eq. (\ref%
{familia_r}) determines the bound 
\begin{equation}
r_{b}(t)=r_{a}(t-r_{a}/c)<r_{a}(t).  \label{bi_monot}
\end{equation}%
For the complementary interval before the next collision$\ (t_{C}>t>t_{0})$,
we have $r_{a}<r_{b}\ $. Because of inequality (\ref{bi_monot}), when the
largest radius $r_{a}(t)$ goes to zero, $r_{b}(t)$ must go to zero as well,
such that this largest radius becomes the natural control parameter for the
dynamics in the neighborhood of the collision. Our next Lemma shows that
velocities $v_{2a},v_{1},$ and $v_{2b}$ must all tend to the speed of light
in modulus when this largest radius $r_{a}$ goes to zero.

\ \textit{Lemma:}

Assuming that a continuous solution $\left( r\left( t\right) ,v\left(
t\right) \right) $ exists in an open neighborhood of \ the collision point $%
r_{a}=0$ and $t=-t_{C}\;$, then we must have that both velocities go to the
speed of light at $t=-t_{C}$

\bigskip \textit{Proof by contradiction:}

Dividing Eq. (\ref{eq1}) by Eq. (\ref{eq8}) we obtain the following equation
for the evolution of $v_{1}$ with $r_{b}$%
\begin{equation}
\frac{dv_{1}}{dr_{b}}=-\frac{\left( 1-v_{2b}\right) }{2\ \left(
v_{1}-v_{2b}\right) }\left\{ \frac{\left( 1-v_{2a}\right) }{r_{a}^{2}\left(
1+v_{2a}\right) }+\frac{\left( 1+v_{2b}\right) }{r_{b}^{2}\left(
1-v_{2b}\right) }\right\} \left( 1-v_{1}^{2}\right) ^{3/2}.  \label{dv_drb}
\end{equation}%
If neither of the velocities goes to the speed of light at the collision,
then by Eq. (\ref{defretadv}) it must be that $\ $%
\begin{eqnarray}
v_{1} &=&v^{C}<1,  \label{bounds} \\
v_{2a} &=&-v^{C},\;  \notag \\
v_{2b} &=&-v^{C}.  \notag
\end{eqnarray}%
Given that the velocities are bounded as of Eq. (\ref{bounds}) and because
of \ Eq.(\ref{bi_monot}), the second term on the right-hand side of \ Eq. (%
\ref{dv_drb}) dominates and integration of this dominant term yields 
\begin{equation}
v_{1}=\frac{k}{r_{b}}+\ ...,  \label{absurd}
\end{equation}%
with $k$ a nonzero constant. Equation (\ref{absurd}) predicts that $v_{1}$
becomes infinite as $r_{b}$ goes to zero, an absurd, as we have assumed that 
$v_{1}=v_{1}^{C}<1.\;$We conclude that $v_{1}\left( r_{a}=0\right) =1$ and $%
v_{2b}\left( r_{b}=0\right) =$ $v_{2a}\left( r_{a}=0\right) =-1$ .

\section{The time transformation}

\bigskip

To motivate our regularizing time-transformation, we assume that at least
one regular orbit exists and construct its formal series expansion in the
neighborhood of the collision. For this we develop the function $1+v_{2a}$
in a power series of $r_{a}$ and the function $1+v_{2b}$ in a power series
of $r_{b}$%
\begin{eqnarray}
1+v_{2a} &\sim &ar_{a}^{q}+a_{1}r_{a}^{q+1}+...  \notag \\
1+v_{2b} &\sim &Br_{b}^{s}+B_{1}r_{b}^{s+1}+...\   \label{catseries}
\end{eqnarray}%
where $s$ and $q$ must be positive because the velocities have a bounded
modulus ($|v_{2}|<$ $1$ ) and $a\ $and$\ B>0\ $are to be determined later.
It is easy to verify that the evolution parameter $u\ $defined as 
\begin{equation}
dt_{1}=du\left( 1+v_{2a}\right) \left( 1-v_{2b}\right) ,  \label{parameter_u}
\end{equation}%
$\ $regularizes Eqs. (\ref{neeq2})-(\ref{neeq3}) and Eqs. (\ref{eq7})-(\ref%
{eq8}) at the collision. The only problematic regularization left is the
right-hand side of \ Eq. (\ref{eq1}), which involves two indefinite limits
at the collision. In the following we show that the second term on the
right-hand side of Eq. (\ref{eq1}) vanishes at the collision. For this we
divide Eqs. (\ref{eq1}) and$\;$(\ref{eq8}) by Eq. (\ref{eq7}) and obtain
differential equations for $v_{1\text{ }}$and $r_{b}$ in terms of the
evolution parameter$\ r_{a}$ 
\begin{equation}
\frac{dv_{1}}{dr_{a}}=-\frac{\left( 1-v_{1}^{2}\right) ^{3/2}}{2\
r_{a}^{2}\left( v_{1}-v_{2a}\right) }\left\{ \left( 1-v_{2a}\right) +\frac{%
\left( 1+v_{2b}\right) \left( 1+v_{2a}\right) r_{a}^{2}}{r_{b}^{2}\left(
1-v_{2b}\right) }\right\}  \label{difterm}
\end{equation}%
\begin{equation}
\frac{dr_{b}}{dr_{a}}=(\frac{v_{1}-v_{2b}}{1-v_{2b}})(\frac{1+v_{2a}}{%
v_{1}-v_{2a}})\sim \frac{1+v_{2a}}{2}+\left( \frac{1+v_{2a}}{2}\right)
^{2}\sim \frac{a}{2}r_{a}^{q}+....\   \label{drb_dra}
\end{equation}%
Eqs. (\ref{difterm}) and (\ref{drb_dra})$\ $\ are converted into ordinary
differential equations by use of the series of \ Eqs. (\ref{catseries}).
Integrating the leading term of Eq.$\ $(\ref{drb_dra})$\ $we find%
\begin{equation}
r_{b}\sim \frac{ar_{a}^{q+1}}{2\left( q+1\right) }.  \label{rb_ra}
\end{equation}%
To obtain a series for $v_{1}$ from \ Eq. (\ref{difterm}), we start by
noticing that the first term on the right-hand side of Eq. (\ref{difterm})
is approximately $2$ in the neighborhood of $r_{a}=0$ 
\begin{equation}
\ \left( 1-v_{2a}\right) \sim 2.  \label{termos2}
\end{equation}%
By use of Eqs. (\ref{catseries}) and (\ref{rb_ra}) we can show that the
second term of the right-hand side of Eq. (\ref{difterm}) is proportional to 
$r_{a}^{qs+s-q}$%
\begin{equation}
\frac{\left( 1+v_{2b}\right) \left( 1+v_{2a}\right) r_{a}^{2}}{%
r_{b}^{2}\left( 1-v_{2b}\right) }\varpropto r_{a}^{qs+s-q}\varpropto \frac{%
(1+v_{2a)}}{(1+v_{2b})}.  \label{secondterm}
\end{equation}%
As the velocity is monotonic, the limit of the ratio $\frac{(1+v_{2a)}}{%
(1+v_{2b})}$ in Eq. (\ref{secondterm}) must either be zero or at the worst
this limit can be a constant value, which implies that $qs+s-q\geq 0$. In
the following we assume that Eq.(\ref{secondterm}) vanishes near the
collision, and obtain a leading approximation to the equations of motion.
This leading approximation in turn calculates $qs+s-q=3$, showing that the
assumption is consistent. The pathological option $qs+s-q=0$ must be
analyzed separately, as then Eq. (\ref{secondterm}) has a finite limit. This
analysis again determines that $qs+s-q=3$, showing that a vanishing limit
for Eq. (\ref{secondterm}) is the only consistent choice.

With the above in mind, the leading terms of Eqs. (\ref{eq1}), (\ref{eq7})
and (\ref{eq8}) in the neighborhood\ of $u=0$ \ are 
\begin{eqnarray}
\frac{dr_{a}}{du} &\sim &4,  \label{dv_ds} \\
\frac{dv_{1}}{du} &\sim &-\frac{4\sqrt{2}}{r_{a}^{2}}\left( 1-v_{1}\right)
^{3/2},  \notag \\
\frac{dr_{b}}{du} &\sim &2\left( 1+v_{2a}\right) .  \notag
\end{eqnarray}%
Choosing $u=0$ at $t=-t_{C}$ and using Eq. (\ref{catseries}) to eliminate $%
v_{2a}$,$\ $Eqs.(\ref{dv_ds}) \ can be integrated, yielding$\ $%
\begin{eqnarray}
r_{a} &\sim &4u,  \notag \\
1+v_{2a} &\sim &au^{q}  \notag \\
1-v_{1} &\sim &32u^{2},  \notag \\
r_{b} &\sim &\frac{2a}{\left( q+1\right) }u^{q+1}.  \label{leadingbunch}
\end{eqnarray}%
Eqs. (\ref{leadingbunch}) and (\ref{catseries}) predict also the leading
dependence of $v_{2b}$ with $u$%
\begin{equation}
1+v_{2b}\sim Br_{b}^{s}=\frac{2^{s}a^{s}B}{\left( q+1\right) ^{s}}u^{s\left(
q+1\right) }.  \label{v2b_s}
\end{equation}%
The last consistency condition on the solution is that $x_{2a}\left(
t\right) $ and $x_{2b}\left( t\right) $\ must describe the same orbit$.\ $%
This is accomplished if there exists a shift function $\Delta u\left(
u\right) >0,$ such that 
\begin{eqnarray}
t_{1}\left( u+\Delta u\left( u\right) \right) &=&t_{2a}\left( u\right) , 
\notag \\
1-v_{1}\left( u+\Delta u\left( u\right) \right) &=&1+v_{2a}\left( u\right) ,
\notag \\
1+v_{2b}\left( u+\Delta u\left( u\right) \right) &=&1-v_{1}\left( u\right) .
\label{merge}
\end{eqnarray}%
$\ $ Using the approximations of Eq. (\ref{leadingbunch})$\ $and (\ref{v2b_s}%
) we can solve Eq. (\ref{merge}) at the leading order of approximation with
a term for $\Delta u\left( u\right) $ given by\ 
\begin{equation}
\Delta u\left( u\right) =2^{1/4}u^{\frac{1}{2}},  \label{citeu}
\end{equation}%
and also $B=$ $s=2,$ $q=1$ and $a=2\sqrt{2}.$ Finally we can express the
velocities in terms of the radii using Eqs. (\ref{leadingbunch}) and (\ref%
{v2b_s}) 
\begin{eqnarray}
1+v_{2a} &\sim &2\sqrt{2}r_{a}+...\;,  \notag \\
1-v_{1} &\sim &2r_{a}^{2}+...,\ \   \notag \\
1-v_{1} &\sim &2\sqrt{2}r_{b}+...,  \notag \\
1+v_{2b} &\sim &2r_{b}^{2}+....  \label{v1series}
\end{eqnarray}

For later use, it is interesting to obtain a further term of the series for $%
v_{1}$, (Eq. (\ref{v1series})) in the following way: \ By use of the leading
terms\ of Eqs. (\ref{leadingbunch}), we can express Eq. (\ref{difterm}) as 
\begin{equation}
\frac{dv_{1}}{dr_{a}}=-\frac{\left( 1-v_{1}^{2}\right) ^{3/2}}{%
r_{a}^{2}\left( 1+v_{1}\right) }+O\left( r_{a}^{4}\right) .  \label{dv_dr}
\end{equation}%
The solution of Eq. (\ref{dv_dr}) with the condition that $v_{1}=1$ at $%
r_{a}=0$ is 
\begin{equation}
v_{1}\left( r_{a}\right) =\frac{\left( 1+Cr_{a}\right) ^{2}-r_{a}^{2}}{%
\left( 1+Cr_{a}\right) ^{2}+r_{a}^{2}}+O\left( r_{a}^{5}\right) ,
\label{int_lips}
\end{equation}%
where $C$ is an integration constant to be determined later. Expanding Eq. (%
\ref{int_lips}) in powers of $r_{a}$ we obtain 
\begin{equation}
1-v_{1}\sim 2r_{a}^{2}+4Cr_{a}^{3}+...,\   \label{extrav}
\end{equation}%
which exhibits the next term in the series for $v_{1}$ of Eq. (\ref{v1series}%
). \ Eq. (\ref{extrav}) also prescribes the next term in the series for $%
v_{2b}$ of Eq. (\ref{v1series}), which can be obtained by replacing $r_{a}$
by $r_{b}$ and $v_{1}$ by $-v_{2b}$ in Eq. (\ref{extrav})%
\begin{equation}
1+v_{2b}\sim 2r_{b}^{2}+4Cr_{b}^{3}.  \label{extravb}
\end{equation}%
This symmetry along time-reversible orbits is illustrated in Figure 3 (i.e.,
that $v_{2b}(r_{b})=-v_{1}(r_{a}=r_{b})$ ). The constant $C$ is related to
the energy $E$ of the orbit and is calculated in the next section. Last, it
is of interest to notice that the term we disregarded in the approximation
below Eq. (\ref{termos2}), corresponding to the information from the past in
Eq. (\ref{difterm}), contributes to the expansion of $\;v_{1}\ $only at $5th$
order in $r_{a}.$

\bigskip

\section{ Two finite variables defined by the energy}

The conserved energy of the Kepler problem is simple and well known, while
the corresponding energy for our relativistic problem is somewhat
unfamiliar. The Poincar\'{e} invariance of the Fokker Lagrangian determines
a four-vector constant of motion, which involves an integral over a
light-cone of the orbit \cite{Fey-Whe,Anderson,Klimenko}. For the
one-dimensional symmetric motion of equal masses, as explained in Ref. \cite%
{Klimenko}, the total energy $E_{T}^{WF}\ \equiv E_{1}+E_{2}=2E\ $can be
simplified in two independent constants (a time-reversed pair, $%
E_{1}=E_{2}=E $)%
\begin{eqnarray}
E_{1} &=&\frac{1}{\sqrt{1-v_{1}^{2}}}-\frac{1}{2r_{a}}-\frac{Y_{a}}{\left(
1+v_{1}\right) },  \label{Ener_1} \\
E_{2} &=&\frac{1}{\sqrt{1-v_{1}^{2}}}-\frac{1}{2r_{b}}+\frac{Y_{b}}{\left(
1-v_{1}\right) },  \label{Ener_2}
\end{eqnarray}%
$\allowbreak $ with $Y_{a}$ and $Y_{b}\ $given by 
\begin{eqnarray}
Y_{a} &\equiv &\frac{\left( 1+v_{1}\right) }{2}%
\int_{t_{1}}^{t_{2a}=t_{1}+r_{a}/c}dt_{1}^{\prime }\frac{v_{1}}{r_{-}^{2}}%
\frac{\left( 1+v_{2}^{-}\right) }{\left( 1-v_{2}^{-}\right) },  \label{Ya} \\
Y_{b} &\equiv &\frac{\left( 1-v_{1}\right) }{2}%
\int_{t_{2b}=t_{1}-r_{b}/c}^{t_{1}}dt_{1}^{\prime }\frac{v_{1}}{r_{+}^{2}}%
\frac{\left( 1-v_{2}^{+}\right) }{\left( 1+v_{2}^{+}\right) },  \label{Yb}
\end{eqnarray}%
where $r_{+},v_{2}^{+}\ $and $r_{-},v_{2}^{-}\ $stand for the radius and
velocity at the advanced and retarded light-cones of particle 1
respectively, as illustrated in Figure 4. The total linear momentum of a
symmetric orbit can also be expressed as $P_{T}^{WF}=P_{1}+P_{2}=0\;$\cite%
{Klimenko}, with the time-reversed pair of constants $P_{1}$ and $P_{2}$
given by%
\begin{eqnarray}
P_{1} &=&-\frac{v_{1}}{\sqrt{1-v_{1}^{2}}}-\frac{1}{2r_{b}}-\frac{1}{2}%
\int_{0}^{t_{2b}}\frac{dt}{r_{+}^{2}}\frac{\left( 1-v_{2+}\right) }{\left(
1+v_{2+}\right) }-\frac{1}{2}\int_{o}^{t_{1}}\frac{dt}{r_{+}^{2}}\frac{%
\left( 1-v_{2+}\right) }{\left( 1+v_{2+}\right) },  \label{Pa_1} \\
P_{2} &=&-P_{1}=\frac{v_{2}}{\sqrt{1-v_{2}^{2}}}+\frac{1}{2r_{a}}-\frac{1}{2}%
\int_{0}^{t_{1}}\frac{dt}{r_{-}^{2}}\frac{\left( 1+v_{1-}\right) }{\left(
1-v_{1-}\right) }-\frac{1}{2}\int_{o}^{t_{2a}}\frac{dt}{r_{-}^{2}}\frac{%
\left( 1+v_{1-}\right) }{\left( 1-v_{1-}\right) },  \label{Pb_1}
\end{eqnarray}%
where $v_{2}\left( t_{1}\right) =-v_{1}\left( t_{1}\right) $ and $v_{1}^{-}$
and $v_{1}^{+}$ stand for the velocity at the retarded and advanced
light-cones respectively. Using the equation of motion (Eq. (\ref{eq1})),
one can check that the time derivative of either Eq.(\ref{Ener_1}) or Eq. (%
\ref{Pa_1})\ \ vanishes. This implies that $\frac{dP_{1}}{dt_{1}}=\frac{dE}{%
dt_{1}}=0$ along the motion. The same applies to Eq. (\ref{Ener_2}), which
is the time-reversed of Eq.(\ref{Ener_1}), and also to Eq. (\ref{Pb_1}), \
the time-reversed of \ Eq. (\ref{Pa_1}). Notice that the energy Eqs.(\ref%
{Ener_1}) and (\ \ref{Ener_2}) have the correct Coulombian limit far away
from the collision, as $Y_{a}$ and $Y_{b}$ vanish for large distances. Using
the near-collision behavior of Sec. III one finds that $Y_{a}$ and $Y_{b}$
as defined by Eqs. (\ref{Ya}) and (\ref{Yb}) have finite limits at the
collision, which is the main motivation to introduce these $Y$ variables as
above. Recalling that Eq. (\ref{eq1}) is the only equation left that is not
regularized simply by the time transformation (\ref{parameter_u}), in the
following we use the energy Eqs. (\ref{Ener_1}) and (\ \ref{Ener_2})\ \ to
obtain regular differential equations to replace Eq. (\ref{eq1}).

The value of $Y_{b}\ $as defined by Eq. (\ref{Yb}) can be approximated by
elimination of \ $dt$\ using the dominant term of \ Eq. (\ref{eq1}) 
\begin{equation}
\frac{1}{2}\int_{t_{2b}}^{t_{1}}\frac{v_{1}}{r_{+}^{2}}\frac{\left(
1-v_{2}^{+}\right) }{\left( 1+v_{2}^{+}\right) }dt_{1}^{\prime
}=-\int_{-v_{2b}}^{v_{1}}\frac{v_{1}}{\left( 1-v_{1}^{2}\right) ^{3/2}}%
dv_{1}+O\left( r_{a}^{5}\right) .  \label{apYb}
\end{equation}%
Evaluating the integral on the right-hand side of \ Eq. (\ref{apYb}) yields
an approximation for $Y_{b}$ near the collision 
\begin{equation}
Y_{b}\simeq -\frac{\sqrt{1-v_{1}}}{\sqrt{1+v_{1}}}+\frac{\left(
1-v_{1}\right) }{\sqrt{1-v_{2b}^{2}}},  \label{Yb_aprox}
\end{equation}%
a finite expression by use of Eq. (\ref{v1series}). Substituting this
expression into the energy Eq. (\ref{Ener_2}) yields the approximation%
\begin{equation}
E\simeq -\frac{1}{2r_{b}}+\frac{1}{\sqrt{1-v_{2b}^{2}}},
\label{Ener_b_aprox}
\end{equation}%
which predicts the same behavior for $v_{2b}\left( r_{b}\right) \ $as \ Eq.(%
\ref{v1series})$.$ Also the near collision behavior of $Y_{a}$ can be
derived from Eq. (\ref{v1series}) 
\begin{equation}
Y_{a}\simeq \frac{\left( 1+v_{1}\right) \left( r_{a}-r_{b}\right) }{2}\simeq 
\frac{\left( 1+v_{1}\right) \sqrt{1-v_{1}}}{2\sqrt{2}}  \label{Ya_aprox}
\end{equation}%
At this point it is interesting to reverse the above argument; noticing that
we could have derived the series for $v_{1}$ ( Eq.(\ref{v1series})) from Eq.
(\ref{Yb_aprox}) with the assumption that $Y_{b}$ is finite. In Sec. III we
had to use the existence of the shift function (Eq. (\ref{citeu})) to
accomplish this same result. The fact that Eq. (\ref{v1series}) can be
obtained from the more elegant hypothesis that $Y_{b}$ is finite all along
the orbit suggests the natural mathematical space for a regular solution;
the set of orbits with finite values for $Y_{a}$ and $Y_{b}$ . Using Eq. (%
\ref{Ener_2}) and the approximations of Eqs. (\ref{v1series}), the constant $%
C\;$entering into Eq. (\ref{int_lips}) can be calculated as 
\begin{equation}
C=-2E.  \label{EC}
\end{equation}%
\qquad The evolution of $v_{1}$ with respect to the parameter $u$ can be
calculated from Eq. (\ref{eq1})$\ $\ 
\begin{equation}
\frac{dv_{1}}{du}=-\frac{\sqrt{1-v_{1}^{2}}}{2}\xi ,  \label{eq1_p}
\end{equation}%
$\ $ where $\xi \ $\ is given by 
\begin{equation}
\xi \equiv \left( 1-v_{1}^{2}\right) \left( \frac{\left( 1-v_{2a}\right)
\left( 1-v_{2b}\right) }{r_{a}^{2}}+\frac{\left( 1+v_{2a}\right) \left(
1+v_{2b}\right) }{r_{b}^{2}}\right) .  \label{eqqsi}
\end{equation}%
It can be shown by use of Eqs. (\ref{v1series}) that $\xi $ as defined by
Eq. (\ref{eqqsi}) has a finite limit at the collision. Notice the two
numerically prohibitive features with Eqs. (\ref{eq1_p}) and (\ref{eqqsi}) :
(i) Eq. (\ref{eqqsi}) has indeterminacies of type $\frac{0}{0}$ , and (ii)
Eq. (\ref{eq1_p}) does not evolve starting from the initial condition $%
v_{1}=1$.\ An alternative way to obtain $v_{1}$ is to solve the energy Eqs. (%
\ref{Ener_1}) and (\ref{Ener_2}) for $v_{1}$. As the energy equations
involve the variables $Y_{a}$ and $Y_{b}$, it must be integrated along with
equations for $Y_{a}\ $and $Y_{b}.$ The needed equations of motion for $%
Y_{a}\ $and$\ Y_{b}$ following from Eqs. (\ref{Ya}) and (\ref{Yb})\ are 
\begin{eqnarray}
\frac{dY_{a}}{du} &=&\frac{\left( v_{1}-v_{2a}\right) \left( 1-v_{2b}\right)
\left( 1+v_{1}\right) }{2r_{a}^{2}}-\frac{\xi }{2}\left( \frac{v_{1}}{\left(
1-v_{1}\right) }-Y_{a}\sqrt{\frac{1-v_{1}}{1+v_{1}}}\right) ,  \label{dYa/du}
\\
\frac{dY_{b}}{du} &=&-\frac{\left( v_{1}-v_{2b}\right) \left(
1+v_{2a}\right) \left( 1-v_{1}\right) }{2r_{b}^{2}}+\frac{\xi }{2}\left( 
\frac{v_{1}}{\left( 1+v_{1}\right) }+Y_{b}\sqrt{\frac{1+v_{1}}{1-v_{1}}}%
\right) .  \label{dYb/du}
\end{eqnarray}%
The system formed by Eqs. (\ref{dYa/du}) and (\ref{dYb/du}) still contains
indeterminacies of type $\frac{0}{0}$, so we are not done yet. In Appendix A
we show that the right-hand sides of Eqs. (\ref{dYa/du}) and (\ref{dYb/du})
have finite limits at the collision if: (i) the orbit has a finite energy
and (ii) the variables $Y_{a}$ and $Y_{b}$ as defined by Eqs. (\ref{Ya}) and
(\ref{Yb}) are finite along the collision at $t=-t_{C}$.

\section{$\ $Numerical calculations with the regularized equations}

\bigskip

We use here a numerical method previously developed by us in Ref.\cite{two}
for the repulsive case. This method approximates $v_{2a}$ and $v_{2b}$ with
two power series and integrates the regular equations (\ref{dYa_reg}) and (%
\ref{appI}) together with the regularized versions of Eqs. (\ref{neeq2})-(%
\ref{neeq3}). This predicts future and past histories for particle $2$, $\
x_{2a}(t_{2})$ and $x_{2b}(t_{2})$. \ Next a minimization scheme modifies
the approximation for $v_{2a}$ and $v_{2b}$ to improve the consistency of
the two histories, until the eventual convergence to a consistent history
for particle $2$ is reached. Some observations are in order : (i) In
Appendix A we show that the right-hand sides of Eqs. (\ref{dYa/du}) and (\ref%
{dYb/du}) have finite limits along a symmetric non-runaway orbit, henceforth
called \ $T_{a}^{f}$\ \ and $T_{b}^{f}$, respectively. The approximation for 
$v_{2a}$ near the collision must be postulated such that that these limits
are satisfied (which is expressed by Eq. (\ref{exprealpha}) of Appendix A ).
(ii) To describe a symmetric orbit in the most economical way, the global
time-reversal symmetry can be embedded in the approximation for the
velocities. According to this time-reversal symmetry, the advanced and
retarded velocities must satisfy 
\begin{eqnarray}
v_{2a}\left( t\right)  &=&-v_{2b}\left( -t\right) ,  \label{novonumero} \\
r_{a}\left( t\right)  &=&r_{b}\left( -t\right) .  \notag
\end{eqnarray}%
To satisfy (i) and (ii) and the near-collision behavior of Eqs.(\ref%
{v1series}) we approximate $v_{2b}$ and $v_{2a}$ with an arbitrary function $%
\theta (r_{a},r_{b})$ such that

\begin{eqnarray}
\frac{1-v_{2a}}{1+v_{2a}} &\equiv &\frac{2}{\left( 3+v_{1}\right) }\sqrt{%
\frac{1+v_{1}}{1-v_{1}}}\frac{r_{a}^{2}}{r_{b}}\theta (r_{a},r_{b}),  \notag
\\
\frac{1+v_{2b}}{1-v_{2b}} &\equiv &\frac{2}{\left( 3-v_{1}\right) }\sqrt{%
\frac{1-v_{1}}{1+v_{1}}}\frac{r_{b}^{2}}{r_{a}}\theta (r_{b},r_{a}).
\label{ansatz}
\end{eqnarray}%
If $\theta (r_{a},r_{b})$ is regular and evaluates to one at the collision,
the above ansatz of Eq. (\ref{ansatz})\textit{\ \ }guarantees that $%
T_{b}^{f} $ of Eq.(\ref{finite}) is \textit{explicitly\ }finite (i.e. there
is no division of zero by zero to spoil the numerical calculations). Notice
that under time reversal $r_{a}$ and $r_{b}$ are interchanged ($%
r_{a}\longleftrightarrow r_{b}$) and $v_{1}$ is exchanged by $-v_{1}$ \ ($%
v_{1}\rightarrow -v_{1}$), such that the two lines of \ Eq. (\ref{ansatz})
are exchanged. This is the embedding of the time-reversal symmetry $%
v_{2a}\longleftrightarrow -v_{2b}$. For numerical convenience, the function $%
\theta (r_{a},r_{b})$ must be postulated in two different ways depending on
the energy of the orbit:

(I) For orbits of $E<<1$, it is convenient to use a rational Pad\'{e} \
approximation defined by%
\begin{equation}
\theta (r_{a},r_{b})\equiv \frac{%
1+K_{1}r_{b}+K_{2}r_{b}^{2}+...+K_{N}r_{b}^{N}}{%
1+k_{1}r_{a}+k_{2}r_{a}^{2}+...k_{N}r_{a}^{N}}.  \label{seriesteta}
\end{equation}%
Eq.(\ref{seriesteta}) is a quotient of a polynomial on $r_{b}$\ over a
polynomial on $r_{a}$, which is constructed for the following reasons: (a)
On the first collision, at $t=-t_{C}$, the $T_{b}^{f}$ (as defined by Eq.(%
\ref{finite}) in Appendix A) is \textit{explicitly\ }finite. (b) On the
second collision, at $t=t_{C}$, the regularization as built with Eq. (\ref%
{seriesteta}) is automatic because the embedded time-reversal symmetry
exchanges $r_{a}$ and $r_{b}$, such that $T_{a}^{f}$ (as defined by Eq. (\ref%
{alsofinite}) in Appendix A), is explicitly finite in the same way.

(II) The atomic energy range, of greatest interest to physics, has $%
E\succsim 1$ , which is a difficult limit for approximation (I). Most of the
counter-intuitive features of shallow energy orbits happen because the
function $\theta $ jumps abruptly from the value of one to another constant
value at the Coulombian limit ($r_{a}\simeq r_{b}\gg 1$). The numerically
correct procedure for shallow energies is to postulate $\theta (v)$ with
Spline interpolation \cite{Akima} on the interval $-1<v_{1}<1$ , defined
such that $\ \theta =1$ at the collision ($v=1$) and such that $T_{a}^{f}$
and $T_{b}^{f}$ are finite at both collisions. This is accomplished with the
definition 
\begin{equation}
\theta \equiv 1+\sqrt{1-v^{2}}P(v).  \label{splines}
\end{equation}%
We use up to 22 intervals to approximate the function $P(v)$ with Splines%
\cite{Akima}. Either the coefficients $k_{1},...,k_{N}$ and $K_{1},...,K_{N}$
of the Pad\'{e} approximation or the polynomial coefficients $c_{i\text{ }}$%
of the cubic Splines are to be determined by the self-consistent
minimization in each case. After these coefficients are substituted into
Eqs. (\ref{ansatz}), and then into Eqs. (\ref{dYa_reg}) and (\ref{appI}) our
regular equations of motion become ordinary differential equations that are
integrated with a standard 9/8 explicit \textit{Runge-Kutta\ }pair,
generating the future and past of particle $2$. Our self-consistent method
calculates two functions that should vanish along a symmetric orbit

\begin{eqnarray}
S_{1}\left( k,t\right) &=&x_{1}\left( t\right) +x_{2a}\left( t\right) ,
\label{marq} \\
S_{2}\left( k,t\right) &=&x_{1}\left( t\right) +x_{2b}\left( t\right) , 
\notag
\end{eqnarray}%
at about $m$ points along the orbit$\;(m\simeq 400).\ $The interpolation
coefficients (either $k_{i},K_{i}$ or $c_{i}$) are changed by a least-square
minimization algorithm (\textit{Levenberg-Marquardt algorithm)} \cite{Nash}.
Notice that if we could find an analytical solution for the orbit, these
coefficients could be calculated by setting $S_{1,2}\;$to zero. In practice
we determine a numerical zero for $S_{1,2}\ $of size $1\times 10^{-5}$(see
Table 1 ). As discussed below Eq. (\ref{extravb}), only at $O(r_{a}^{5})$
the information of the past becomes important near the collision, such that
we should use $N\geq 5,$ to include past information into Eq. (\ref{difterm}%
).\ 

We are now ready to start the integrations from $r_{a}=r_{b}=0,$ using the
approximate coefficients and $E\;$already calculated. The complete set of
equations includes Eqs. (\ref{dYb/du}) and (\ref{dYa/du}) together with Eqs.
(\ref{neeq2}) and (\ref{neeq3})$\ $in terms of $r_{a},r_{b},Y_{a}$ and $%
Y_{b}.$ The initial condition $r_{a}=0,r_{b}=0,v=1,$ expressed in terms of
the regular variables reads 
\begin{eqnarray}
Y_{a}\left( u=0\right) &=&0,  \label{iniYs} \\
Y_{b}\left( u=0\right) &=&\sqrt{2}.  \notag
\end{eqnarray}%
\ The energy $E$ is a parameter that appears explicitly in the regular
equation of motion, and the numerical procedure fixes $E$ while the
interpolation coefficients are adjusted by the minimization scheme. The
velocity $v_{1}\;$is calculated numerically by solving Eqs. (\ref{Ener_1})
and (\ref{Ener_2}) for $v_{1}.$ The numerically calculated orbits using the
Pad\'{e} approximation of Eq. (\ref{seriesteta}) are shown in Figure 5 for
four different energies ( $E=-1.0$ $,E=0.1,E=0.5$ and $E=0.8$ ). In Table 1
we list the quantities related to these numerically calculated orbits.
Notice in Table 1 that the energy becomes negative when the maximum
light-cone radius is lesser than the classical electronic radius. The orbits
at atomic energies ($E\lesssim 1$) have two clearly separated regions: (i) a
near-collision region where the velocity is very close to one and (ii) the
turning region, where $r_{a}\simeq r_{b}\gg 1$ and $\theta \rightarrow \frac{%
3}{2r}$ . This last segment of such orbits approximates the turning region
of Coulombian orbits. In the collision region a relativistic orbit must
deviate from a Coulombian orbit, because $v_{1}\rightarrow \infty $ on the
collision for Coulombian orbits. The transition between these two regions is
abrupt, as illustrated in Figure 6. In Figure 7 we magnify the region of
discontinuity for various energies, illustrating that the discontinuity in $%
v $ changes shape with increasing $E.$ In Figure 8 we show the numerically
calculated orbits using the Spline interpolation for $\theta (v)$ ( Eq. (\ref%
{splines})). The energy $E\;$and the relative error of these atomic orbits
are shown in Table 2. Notice in Table 2 that the value of $r\theta $ is
converging to $3/2$ as it should for Coulombian orbits.

\section{\protect\bigskip Conclusions and Discussion}

We derived a differential equation that is regular along the collision of
two equal masses with attractive interaction of the action-at-a-distance
electrodynamics, allowing the numerical study of these orbits for the first
time. Our regular numerical method starts the integration exactly from the
collision. Our procedure is not covariant because we restricted the work to
symmetric orbits, having the energy as the only free parameter (the other
parameter should be the Lorentz boost parameter). A covariant treatment
shall be left for future work, along the lines of Appendix B of Ref. \cite%
{two}. The numerical results of Ref. \cite{two} suggest that at least for
the repulsive equal-mass case a boosted symmetric orbit is already the
general nonrunaway solution. The generalization of this numerical study to
the attractive case awaits a covariant regularization.

Some failed attempts taught us that the different-mass attractive case is
much more involved, and possibly not even regularizable. At present we do
not even know how to coin a formal series solution near the collision for
this case. One reason for that is the complexity of the energy expressions
analogous to Eqs.(\ref{Ener_1}) and (\ \ref{Ener_2}) \cite{Klimenko}. This
fact confuses the definition of the finite variables for the different-mass
case and the problem needs further study.

The two-body problem with repulsive interaction also displays a singularity
at high energies, but surprisingly enough this is not because the particles
collide (they never do). The singularity appears in the right-hand side of
Eq. (\ref{repulsemotion}) when the particles come to the speed of light (
the denominator containing ($1-v$) vanishes). \ It was found by several
authors, first in Ref. \cite{KernerHill}\ and later in Ref. \cite{Igor-pair}%
, that the particles reach the speed of light and turn back keeping a \emph{%
minimum distance }of approximation of about one classical electronic radius.
The regularization of this problem shall be published elsewhere \cite%
{PhysRevE}. This minimum distance of closest approach of about one classical
electronic radius is a kind of exclusion principle of the
action-at-a-distance theory. It is of interest to notice that this exclusion
behavior is already found with the post-Galilean low-velocity approximation
to the Fokker Lagrangian, the Darwin Lagrangian\cite{Anderson}; The
algebraic-differential equations of motion of the Darwin Lagrangian were
studied analytically in \ \cite{Radiusofelec}\ and the phenomenon of closest
approximation was discovered. The distance of closest approximation with the
Darwin Lagrangian is found to be exactly the classical electronic radius, as
well as with the Fokker Lagrangian \cite{KernerHill}.

\section{Appendix A : Regularity of the Y-derivatives}

In the following we show that the right-hand sides of Eqs. (\ref{dYa/du})
and (\ref{dYb/du}) both have a finite limit at the collision if: (i) the
orbit has a finite energy and (ii) variables $Y_{a}$ and $Y_{b}$ as defined
by Eqs. (\ref{Ya}) and (\ref{Yb}) are finite along the collision at $%
t=-t_{C} $. It is convenient to introduce the variables \ $z\equiv \sqrt{%
1-v_{1}}$ and $w\equiv \sqrt{1+v_{1}}$ and to define two manifestly finite
quantities$\ \alpha $ and $\beta $ as 
\begin{eqnarray}
\alpha &\equiv &\frac{1+v_{2a}}{\sqrt{1-v_{1}}},  \notag \\
\beta &\equiv &\frac{1+v_{2b}}{(1-v_{1})^{2}}.
\end{eqnarray}%
Notice that $z=0$ and $w=\sqrt{2}$ at \thinspace $v_{1}=1$, and the finite
limiting values of $\alpha $ and $\beta \ $at the collision as predicted by
Eqs. (\ref{v1series}) are : $\alpha =2$ and $\beta =1/4$.\ A concise way to
obtain the expansions of (\ref{v1series}) is by solving the energy Eq. (\ref%
{Ener_1}) for $r_{a}$ and the energy Eq. (\ref{Ener_2} ) for $r_{b}$%
\begin{eqnarray}
r_{a} &=&\frac{w^{2}z}{2\left( -w^{2}zE-zY_{a}+w\right) }\equiv \frac{w\rho
_{a}}{2}z,  \notag \\
r_{b} &=&\frac{z^{2}w}{2\left( -z^{2}wE+wY_{b}+z\right) }\equiv \frac{\rho
_{b}}{2Y_{b}}z^{2}.  \label{rab_zwYab}
\end{eqnarray}%
$\ $\ The variables $\rho _{a}$ and $\rho _{b}$ as defined by Eq. (\ref%
{rab_zwYab}) have finite limiting values at the collision, $\rho _{a}^{0}=1$
and $\rho _{b}^{0}=1$. It is nice to observe that these finite limits are a
consequence of $E$ , $Y_{a}$ and $Y_{b}$ being finite in (\ref{rab_zwYab}).
We arrive then at a concise definition of a regular orbit: one defined by
finite values of $E$ , $Y_{a}$ and $Y_{b}$. The derivative of $Y_{a}$ (Eq. (%
\ref{dYa/du}) ) can be expressed in terms of these finite quantities as

\begin{eqnarray}
\frac{dY_{a}}{du} &=&T_{a}^{f}+z\left( 4\frac{\alpha }{\rho _{a}^{2}}+\frac{1%
}{2w}\xi Y_{a}+2w^{2}\alpha \beta \frac{Y_{b}}{\rho _{b}^{2}}\right) -4z^{2}%
\frac{\beta }{\rho _{a}^{2}}+  \label{dYa_reg} \\
&&+z^{3}\left( 2\alpha \frac{\beta }{\rho _{a}^{2}}-2w^{2}\alpha \beta \frac{%
Y_{b}}{\rho _{b}^{2}}\right) +6z^{4}\frac{\beta }{\rho _{a}^{2}}%
-4z^{5}\alpha \frac{\beta }{\rho _{a}^{2}},  \notag
\end{eqnarray}%
where $\xi $, defined in Eq. (\ref{eqqsi}), has the explicitly finite
expression%
\begin{equation}
\xi =\left( \frac{16\left( 1-\alpha z/2\right) \left( 1-\beta z^{4}\right) }{%
\rho _{a}^{2}}+\frac{4Y_{b}\alpha \beta z^{3}w^{2}}{\rho _{b}^{2}}\right) .
\label{lasthing}
\end{equation}%
At $z=0$ the only nonzero term on the right-hand side of Eq. (\ref{dYa_reg})
is

\bigskip 
\begin{equation}
T_{a}^{f}=4\left( 1-z\left( wE-Y_{a}/w\right) \right) ^{2},
\label{alsofinite}
\end{equation}%
which is finite at the collision ($z=0,w=0$), such that the derivative of $%
Y_{a}$ as written in Eq. (\ref{dYa_reg}) is manifestly finite at the
collision. The $Y_{b}$ derivative can be obtained by substitution of Eq. (%
\ref{rab_zwYab})\ into Eq. (\ref{dYb/du}), and some algebraic manipulations
yield 
\begin{eqnarray}
\frac{dY_{b}}{du} &=&T_{b}^{f}+\left( 16+\frac{1}{2w}\xi +8wE\alpha
Y_{b}^{2}-4w\alpha \frac{Y_{b}}{\rho _{a}^{2}}+\frac{8}{w}\alpha
Y_{a}Y_{b}^{2}\right) +  \label{appI} \\
&&+z\left( \frac{8}{wY_{b}}-8E\alpha Y_{a}Y_{b}^{2}-16wE+2\alpha \frac{%
Y_{b}^{2}}{\rho _{b}^{2}}-4w^{2}E^{2}\alpha Y_{b}^{2}-\frac{4}{w^{2}}\alpha
Y_{a}^{2}Y_{b}^{2}\right)  \notag \\
&&+z^{2}\allowbreak \left( 2w^{3}\alpha \beta \frac{Y_{b}^{2}}{\rho _{b}^{2}}%
-16\frac{E}{Y_{b}}-\frac{1}{2w}\xi \right)  \notag \\
&&+z^{3}\left( 8w\frac{E^{2}}{Y_{b}}-8w\beta \frac{Y_{b}}{\rho _{a}^{2}}%
+2\alpha \beta \frac{Y_{b}^{2}}{\rho _{b}^{2}}\right) +4wz^{4}\allowbreak
\alpha \beta \frac{Y_{b}}{\rho _{a}^{2}},  \notag
\end{eqnarray}%
where

\begin{equation}
T_{b}^{f}=\frac{4Y_{b}}{z}\left( 2w-\alpha Y_{b}\right) .  \label{finite}
\end{equation}%
The term $T_{b}^{f}$ defined by Eq. (\ref{finite}) was singled out in Eq. (%
\ref{appI}) because it contains a division of zero by zero at the collision,
and is the only part of Eq. (\ref{appI}) that could in principle be
singular. To show that $T_{b}^{f}$ has a \textit{finite} limit along an
orbit of \ finite energy, we first use the energy Eq. (\ref{Ener_2}) to
express $z$ as

\begin{equation}
z=\frac{\left( r_{b}+\sqrt{r_{b}}\sqrt{2w^{2}Y_{b}+r_{b}+4w^{2}EY_{b}r_{b}}%
\right) }{w\left( 1+2Er_{b}\right) }.  \label{expreZ}
\end{equation}%
Substituting the approximation of Eq. (\ref{Yb_aprox}) for $Y_{b}$ 
\begin{equation}
Y_{b}\simeq \sqrt{2}-\frac{z}{w}+O\left( z^{4}\right) ,  \label{elvis}
\end{equation}%
into Eq. (\ref{expreZ}), we find the expansion for $z^{2}\left( r_{b}\right)
\equiv \left( 1-v_{1}\right) \ $to be%
\begin{equation}
z^{2}=2r_{b}\sqrt{2}-4\sqrt{2}Er_{b}^{2}+O\left( r_{b}^{3}\right)
\label{z2rb}
\end{equation}%
Last, as discussed above Eq. (\ref{extravb})\ and illustrated in Figure 3,
the particle exchange symmetry along symmetric orbits implies that the
function $v_{1}(r_{b})$ is equal to the function $-v_{2b}(r_{a})$. Changing
the argument of Eq. (\ref{z2rb}) from $r_{b}$ to $r_{a\text{ }}$and
eliminating this $r_{a}$ in favor of $v_{1}$ with Eq. (\ref{v1series}) we
obtain 
\begin{equation}
\alpha \equiv \frac{1+v_{2a}}{z}=2-2\sqrt{2}E\ z+O\left( z^{2}\right)
\label{exprealpha}
\end{equation}%
Substituting Eq. (\ref{exprealpha}) for $\alpha $ into the formula for $%
T_{b}^{f}$ , Eq. (\ref{finite}), and expanding we can determine that $%
T_{b}^{f}\ $\ is \textit{finite} and nonzero at the collision.

\section{Acknowledgments:}

\bigskip E. B. Hollander acknowledges a Fapesp scholarship, proc. 99/08316-8
and J. De Luca acknowledges the partial support of CNPQ, Brazil.

\section{\protect\bigskip Figure Captions}

Fig. 1 A symmetric trajectory, arbitrary units. Indicated are the maximum
light-cone distance $r_{0}$, the velocity $v_{1}$ of particle $1$ at time $%
t_{1}$ and the corresponding retarded and advanced velocities of particle $2$%
, $v_{2b}$ and $v_{2a}$. The trajectories of particle $1$ (solid line on the
right-hand side) and particle $2$ (solid line on the left-hand side) are
illustrated from the outgoing collision at $t=-t_{C}$ until the ingoing
collision at $t=t_{C}.$

\bigskip

Fig. 2 The orbits of particle $1$ (solid line on the right-hand side) and
particle $2$ (solid line on the left-hand side) in arbitrary units.
Indicated is the symmetric point $v_{1}=0$ at $t=0$. Also shown are the the
velocity $v_{1}$ of particle $1$ at time $t_{1}$ and the corresponding
retarded and advanced velocities $v_{2b}$ and $v_{2a}$ of particle $2.$

\bigskip

Fig. 3 A symmetric orbit in arbitrary units. The symmetric orbit is
completely defined by only two functions $r(t)$ and $v(t)$ , from which the
quantities $r_{a},$ $r_{b},$ $v_{1}$, $v_{2b\text{ }}$and $v_{2a}$ are
determined. Notice that $-v_{2a}$ when $r_{b}=r$ is the same as $v_{1}$ when 
$r_{a}=r$.

\bigskip

Fig. 4 A symmetric orbit in arbitrary units. Illustrated are the segments of
trajectory relevant to the evaluation of $Y_{a}$ and $Y_{b}$ and the
quantities $r^{+},r^{-},$ $v_{2}^{+}$ and $v_{2}^{-}$.

\bigskip

Fig. 5 Numerically calculated trajectories using the Pad\'{e} approximation,
in units where $c=e=m=1$. For each value of the energy $E$, the regularized
integrator starts from the initial condition $r_{a}=r_{b}=0$, $v_{1}=1$, $%
Y_{a}=0$ and $Y_{b}=\sqrt{2}$ $.$ The four different orbits shown have
energies $E=-1.0$, $0.1$, $0.5$ and $0.8$ as indicated.

\bigskip

Fig. 6. Numerically calculated $\theta (v_{1})$ (as defined by Eq. (\ref%
{seriesteta}) ) plotted versus the velocity $v_{1}$, in units where $c=e=m=1$%
. This figure illustrates the jump in $\theta $ near $v=1$ at $E=0.8$ (Dash
Dot), $E=0.9$ (Dot) , $E=0.95$ (Dash) and $E=0.99$ (solid line).

\bigskip

\bigskip Fig. 7. Numerically calculated $\theta (v_{1})$ (as defined by Eq. (%
\ref{seriesteta}) ) plotted versus the velocity $v_{1}$, in units where $%
c=e=m=1$. In this figure the discontinuity region near $v=1$ is blown-up.
Energies are $E=0.8$ (Dash Dot), $E=0.9$ (Dot) , $E=0.95$ (Dash) and $E=0.99$
(solid line). Notice that the jump in $\theta (v)$ at $E=0.99$ is quite
differently from the jump at the other energies.

\bigskip

Fig. 8 Numerically calculated trajectories using the Spline interpolation,
in units where $c=e=m=1$. For each value of the energy $E$, the regularized
integrator starts from the initial condition $r_{a}=r_{b}=0$, $v_{1}=1$, $%
Y_{a}=0$ and $Y_{b}=\sqrt{2}$ $.$ The four different orbits have energies $%
E=0.8$, $0.9$, $0.95$ and $0.99$ as indicated.

\bigskip

Table 1: Numerically calculated orbits using the Pad\'{e} approximation.
Indicated are the energy $E$, the maximum light-cone distance $r_{0}$ and
the relative size of the first and last Pad\'{e} coefficients, as well as
the relative error of the minimization scheme, $\Delta x/r_{0}$. Notice that
at the negative energy $E=-1.0$ the maximum radius $r_{0}$ is less than the
classical electronic radius $r^{\ast }\equiv \frac{e^{2}}{mc^{2}}=1$.

\bigskip

Table 2: \ Numerically calculated orbits using the Spline interpolation.
Indicated are the energy $E$, the maximum light-cone distance $r_{0}$, the
saturation value $r_{0}\theta (v=0)$ and the relative error of the
minimization scheme, $\Delta x/r_{0}$. Notice again that at $E=-1.0$ the
maximum radius $r_{0}$ is less than the classical electronic radius $r^{\ast
}\equiv \frac{e^{2}}{mc^{2}}=1$ and that $r_{0}\theta (v=0)$ approximates
the Coulombian limiting value of $3/2$ as $E$ tends to $1.$

\end{document}